Relativistic particle-like structures associated with multi-soliton solutions of (1+2)-dimensional Sine-Gordon equation

Yair Zarmi
The Jacob Blaustein Institutes for Desert Research
Ben-Gurion University of the Negev
Midreshet Ben-Gurion, 84990, Israel

**Abstract**
The Sine-Gordon equation in (1+2) dimensions has $N$-soliton solutions that propagate at velocities that are lower than the speed of light ($c = 1$), for any $N \geq 1$. A first integral of the equation, which vanishes identically on the single-soliton solution, maps multi-soliton solutions onto structures that are localized around soliton junctions. The profile of such a structure obeys the (1+2)-dimensional linear wave equation, driven by a source term, which is constructed from a multi-soliton solution of the Sine-Gordon equation. If the localized solutions of the source-driven wave equation are interpreted as mass densities, they emulate free, spatially extended, massive relativistic particles. This physical picture is summarized in terms of a Lagrangian density for a dynamical system, in which the Sine-Gordon equation and the linear wave equation are coupled by a small coupling term. The Euler-Lagrange equations of motion allow for solutions, which, in lowest order in the coupling constant are the soliton solutions of the Sine-Gordon equation, and the first-order component are the structures that emulate spatially extended relativistic particles.

## 1. Introduction

Soliton solutions of integrable nonlinear evolution equations in (1+1) dimensions are localized in position at any fixed time. Hence, in some applications to classical systems, solitons have been viewed as emulation of spatially extended but localized particles with finite masses:

$$m = \int_{-\infty}^{+\infty} u(t,x) dx \quad , \tag{1}$$

where $u(t,x)$ is a soliton solution. Examples of this approach are the dynamics of solitons subjected to external forces in, e.g., [1, 2]), the peakon (a localized structure) [3], the compacton, (a solitray wave solution of finite support) [4] and localized static solitons - in KdV-type systems [5].

Localized solutions, for which

$$\int_{-\infty}^{+\infty} \left| u(t,x)^2 \right| dx < \infty \quad , \tag{2}$$

have been sometimes viewed as candidates for normalizable wave functions, which could be used as starting points for the construction of bound states within the framework of quantum-field theoretical models. This idea has been discussed extensively in the High Energy literature (see, e.g., [6-16].) In particular, the connection between soliton solutions of the (1+1)-dimensional Sine-Gordon equation and particles within a field-theoretic context has been studied (see, e.g., [14]).

Line-solitons soliton solutions in higher space dimensions cannot be exploited for the goals discussed above because they are not localized in space. This has led to the extensive search for integrable higher-dimensional nonlinear evolution equations that have spatially localized solutions. Prime examples of such equations that describe classical systems are the Kadomtsev-Petviashvili I equation [17], which has "lump" solutions [17, 18], the Davey-Stewartson equation [19, 20], the Gardner equation [21, 22] and the Nizhnik-Veselov-Novikov system [23-25]. The search for such equations in Quantum-Field Theory began with the discovery of the 't Hooft-Polyakov monopole [6, 7] - a spatially localized solution of the (1+3)-dimensional nonlinear Klein-Gordon equation.

The approach discussed above has focused on nonlinear evolution equations that have spatially localized solutions. In this paper, an alternative is presented: The generation of localized structures, which emulate spatially extended particles, from multi-line soliton solutions in more than one space dimension. The (1+2)-dimensional Sine-Gordon equation is considered, as it offers an interesting possibility. It has slower-than-light multi-soliton solutions. If the spatially localized structures generated from these solutions are interpreted as mass densities, then they emulate free, spatially extended relativistic particles.

The localized structures are generated by a functional of the solution, which, as a direct consequence of the Sine-Gordon equation, vanishes on the single-soliton solution. This functional maps multi-soliton solutions onto structures, which are localized around soliton junctions and obey the linear wave equation, driven by a localized source term. The latter is constructed from a multi-soliton solution of the Sine-Gordon equation. This physical picture can be cast in the form of the Euler-Lagrange equations of a (1+2)-dimensional dynamical system, in which the Sine-Gordon equation and the linear wave equations are coupled by a small coupling term.

The construction of solutions of the Sine-Gordon equation in (1+1) and (1+2) dimensions is reviewed in Section 2. The generation of localized structures from multi-soliton solutions is presented in Section 3. The source-driven wave equation for a localized structure and the resulting interpretation of the structure as the mass density of a free, spatially extended relativistic particle are presented in Section 4. The Lagrangian of the coupled Sine-Gordon and linear wave equations is presented in Section 5. The properties of space-like momentum vectors that are pertinent to the results presented here are discussed in Appendix I. Section 6 offers concluding comments.

**2. Soliton solutions of the Sine-Gordon equation**
**2.1 Review of (1+1) dimensional case**
The Sine-Gordon equation [26-34],

$$\partial_\mu \partial^\mu u + \sin u = 0 \quad , \tag{3}$$

is integrable in (1+1) dimensions [35]. The Hirota algorithm [36] for the construction of its soliton solutions is based on a transformation, originally proposed in the cases of one- and two-soliton solutions [32,33]:

$$u(x;Q) = 4\tan^{-1}\left[g(x;Q)/f(x;Q)\right] . \tag{4}$$

In Eq. (4),

$$Q \equiv \left\{q^{(1)}, q^{(2)}, ..., q^{(N)}\right\} , \tag{5}$$

and $x$ and $q$ are, respectively, (1+1)-dimensional coordinate and momentum vectors. In addition:

$$g(x;Q) = \sum_{\substack{1 \le n \le N \\ n \text{ odd}}} \left( \sum_{1 \le i_1 < \cdots < i_n \le N} \left\{ \prod_{j=1}^{n} \varphi(x; q^{(i_j)}) \prod_{i_l < i_m} V\left(q^{(i_l)}, q^{(i_m)}\right) \right\} \right) , \tag{6}$$

$$f(x;Q) = 1 + \sum_{\substack{2 \le n \le N \\ n \text{ even}}} \left( \sum_{1 \le i_1 < \cdots < i_n \le N} \left\{ \prod_{j=1}^{n} \varphi(x; q^{(i_j)}) \prod_{i_l < i_m} V\left(q^{(i_l)}, q^{(i_m)}\right) \right\} \right) , \tag{7}$$

$$\varphi(x; q^{(i)}) = e^{q_\mu^{(i)} x^\mu + \delta^{(i)}} , \tag{8}$$

where $\delta^{(i)}$ is a constant free phase,

$$q_\mu^{(i)} q^{(i)\mu} = -1 , \tag{9}$$

and

$$V(q, q') = \frac{1 + q_\mu q'^\mu}{1 - q_\mu q'^\mu} . \tag{10}$$

The solitons show up in the current density:

$$J_\mu = \partial_\mu u(x) . \tag{11}$$

Finally, the $N$-soliton solution is a Lorentz scalar, as it is constructed in terms of Lorentz scalars ($x \cdot q^{(i)}$, $q^{(i)} \cdot q^{(j)}$, $1 \le i, j \le N$). (This was first observed in the case of the single-soliton solution [33].) If $x$ and $q^{(i)}$, $1 \le i \le N$ in two reference frames are connected by a Lorentz transformation, $L$, then:

$$u\left(x, \left\{q^{(1)}, ..., q^{(N)}\right\}\right) = u\left(\tilde{x}, \left\{\tilde{q}^{(1)}, ..., \tilde{q}^{(N)}\right\}\right) \quad \left(x = L \cdot \tilde{x}, \quad q^{(i)} = L \cdot \tilde{q}^{(i)} , \quad 1 \le i \le N\right) . \tag{12}$$

## 2.2 Higher space dimensions

The attempt to extend the Hirota algorithm to the (1+2)-dimensional Sine-Gordon equation produced one- and two-soliton solutions, but encountered an obstacle in the case of three (let alone more than three) solitons [37]. For a three-soliton solution to exist, one of the three momentum vectors, from which the solution is constructed via Eqs. (4)-(10)), had to be a linear combination of the other two momenta. In later years, it was shown that the (1+2)-dimensional Sine-Gordon equation does not pass integrability tests employed in nonlinear dynamics [38-41].

The obstacle exposed by Hirota is the key to the extension of his algorithm to $N$-soliton solutions in (1+2) dimensions for any $N \geq 3$ [43]. For a solution with $N \geq 3$ solitons to exist, $(N-2)$ of the momentum vectors in Eqs. (4)-(10) must be linear combinations of just two of them:

$$q^{(i)} = \alpha_i q^{(1)} + \beta_i q^{(2)} \quad (3 \leq i \leq N) \ . \tag{13}$$

Owing to Eq. (13), a multi-soliton (1+2)-dimensional solution propagates as a whole at one velocity, $\vec{v}$. The solutions are divided into two subsets: Solutions with $|\vec{v}| \geq c = 1$, and solutions with $|\vec{v}| < c$. This paper focuses on the latter subset.

The Hirota algorithm generates soliton solutions also for the (1+3)-dimensional Sine-Gordon equation. The one-and two-soliton solutions are mere spatially rotated solutions in (1+2) dimensions. For a solution with $N \geq 3$ solitons to exist, every triplet of vectors used in the construction the solution must obey the constraint [42]:

$$\begin{vmatrix} q_1^{(1)} & q_2^{(1)} & q_2^{(1)} \\ q_1^{(2)} & q_2^{(2)} & q_2^{(2)} \\ q_1^{(3)} & q_2^{(3)} & q_2^{(3)} \end{vmatrix}^2 = \begin{vmatrix} q_1^{(1)} & q_2^{(1)} & q_0^{(1)} \\ q_1^{(2)} & q_2^{(2)} & q_0^{(2)} \\ q_1^{(3)} & q_2^{(3)} & q_0^{(3)} \end{vmatrix}^2 + \begin{vmatrix} q_1^{(1)} & q_3^{(1)} & q_0^{(1)} \\ q_1^{(2)} & q_3^{(2)} & q_0^{(2)} \\ q_1^{(3)} & q_3^{(3)} & q_0^{(3)} \end{vmatrix}^2 + \begin{vmatrix} q_2^{(1)} & q_3^{(1)} & q_0^{(1)} \\ q_2^{(2)} & q_3^{(2)} & q_0^{(2)} \\ q_2^{(3)} & q_3^{(3)} & q_0^{(3)} \end{vmatrix}^2 \ . \tag{14}$$

Eq. (14) offers a rich variety of solutions. These contain a subset of interest for this paper: Solutions with $N \geq 2$ solitons in (1+3) dimensions that propagate as a whole at one velocity, $|\vec{v}| < c$.

For $N = 2$, this is a direct consequence of the discussion in Appendix I. For $N \geq 3$, such solutions are constructed from momentum vectors, for which Eq. (13) holds. Then, Eq. (14) is obeyed trivially. For each triplet of the $N$ vectors, all four determinants in Eq. (14) vanish separately. As a result, multi-soliton solutions in (1+3) dimensions, which propagate rigidly at a velocity $|\vec{v}| < c$, are obtained by applying three-dimensional rotations to such (1+2)-dimensional solutions. Therefore, *this paper focuses on slower-than-light* (1+2)-*dimensional soliton solutions*.

### 2.3 Construction of slower-than-light solutions in (1+2) dimensions [43]

One first constructs *static* (stationary and time independent) solutions through Eqs. (4)-(10). All the momentum vectors are of the form:

$$\tilde{q}^{(i)} = \left\{0, \cos\psi^{(i)}, \sin\psi^{(i)}\right\} \quad , \quad (1 \leq i \leq N) \ . \tag{15}$$

The coefficients of Eq. (10) then obtain the form:

$$V\left(\tilde{q}^{(i)}, \tilde{q}^{(j)}\right) = \left(\tan\left(\frac{\psi^{(i)} - \psi^{(j)}}{2}\right)\right)^2 > 0 \ . \tag{16}$$

Moving multi-soliton solutions, whose velocity is lower than $c$, are Lorentz transforms in (1+2) dimensions of static solutions; Lorentz transforming the coordinate vector, $x$, and the momentum vectors in a static solution yields a moving soliton solution, which is Lorentz invariant.

All solutions obtained as described above propagate rigidly at velocities that are lower than $c = 1$. This characteristic is intimately related to the fact that the scalar product of any two of the momentum vectors, from which the solution is constructed, obeys:

$$\left|q^{(i)} \cdot q^{(j)}\right| < 1 \quad (1 \leq i \neq j \leq N) \ . \tag{17}$$

This procedure is invertible. Given a solution with $N \geq 2$ solitons that is moving rigidly at a velocity, $|\vec{v}| < c$, the momenta obey Eq. (13), and all pairs obey Eq. (17). This ensures that a Lorentz transformation to a rest frame, where the solution is static, exists.

Multi-soliton solutions that propagate at velocities that are equal to, or greater than *c* are of no interest for this paper. They are constructed via Eqs. (4)-(10) with momentum vectors, in which all pairs obey:

$$\left|q^{(i)} \cdot q^{(j)}\right| \geq 1 \quad (1 \leq i \neq j \leq N) \ . \tag{18}$$

**3. Mapping of (1+2)-dimensional multi-soliton solutions onto localized structures**

The first step is the derivation of an identity that is obeyed by the single-soliton solution (in all space dimensions, *n* = 1–3). The *x*-dependence of a single-soliton solution is of the form

$$u(x;q) = h(\xi + \delta) \quad \left(\xi = q \cdot x \ , \ q_\mu q^\mu = -1\right) \ . \tag{19}$$

In Eq. (21), $\delta$ is the constant free phase of Eq. (8). Using Eqs. (19) and (9), the Sine-Gordon equation for a single-soliton solution in any space dimension becomes

$$-h'' + \sin h = 0 \ . \tag{20}$$

Accounting for the vanishing boundary value of the *h* at $\xi \to -\infty$, a first integral is obtained:

$$-\frac{1}{2}(h')^2 + (1 - \cos h) = 0 \ . \tag{21}$$

The relativistically invariant form of Eq. (21) in any moving frame is:

$$R[u] = \frac{1}{2}\partial_\mu u \partial^\mu u + (1 - \cos u) = 0 \ . \tag{22}$$

Eq. (22) is obeyed by any single-soliton solution in (1+*n*) dimensions for *n* =1–3. However, it is violated by all multi-soliton solutions. *R*[*u*] then constitutes a mapping of the solution onto structures that are localized in the vicinity of soliton junctions in *u*. As an example, consider the two-soliton solution. Based on Eqs. (4)-(10), one can rewrite the solution as a function of the arguments of the exponential in Eq. (8):

$$\xi_i = q^{(i)} \cdot x + \delta^{(i)} \quad (i = 1,2) \ . \tag{23}$$

Here $x$ and $q^{(i)}$ are vectors in (1+2) dimensions. Substituting Eqs. (4)-(10) in the definition of $R[u]$, Eq. (22), one obtains:

$$R[u] = \frac{64\, e^{\xi_1 + \xi_2}\, V(q^{(1)}, q^{(2)})}{\left(1 + V(q^{(1)}, q^{(2)})\right)\left\{1 + e^{2\xi_1} + e^{2\xi_2} + e^{2(\xi_1 + \xi_2)}\left(V(q^{(1)}, q^{(2)})\right)^2 + 2 e^{\xi_1 + \xi_2}\left(1 + V(q^{(1)}, q^{(2)})\right)\right\}} \quad . \tag{24}$$

The sign of $V(q^{(1)}, q^{(2)})$ of Eq. (10) determines the properties of $R[u]$. In the case of slower-than-light multi-soliton solutions, thanks to Eq. (17), $V(q^{(1)}, q^{(2)}) > 0$. Consequently, $R[u]$ is positive definite when computed for these solutions. In addition, $R[u]$ then falls off exponentially along each soliton line away from the soliton junction. For example, along soliton no. 1, $\xi_1$ is of $O(1)$, whereas, away from the junction, $|\xi_2|$ is large. $R[u]$ then falls off exponentially as:

$$R[u] \underset{|\xi_2| \to \infty}{\to} \frac{64\, e^{\xi_1}\, V(q^{(1)}, q^{(2)})}{\left(1 + V(q^{(1)}, q^{(2)})\right)\left\{1 + e^{2\xi_1}\left(V(q^{(1)}, q^{(2)})\right)^2\right\}} e^{-|\xi_2|} + O\left(e^{-2|\xi_2|}\right) \quad . \tag{25}$$

Finally, $R[u]$ has a maximum:

$$R\left[u\left(\xi_1 = \xi_2 = \frac{1}{2}\log V(q_1, q_2)\right)\right] = \left(16 V(q^{(1)}, q^{(2)}) \Big/ \left(1 + V(q^{(1)}, q^{(2)})\right)^2\right) > 0 \quad . \tag{26}$$

The image of a multi-soliton solution under $R[u]$ will be called a *vertex map*. A vertex map of a slower-than-light multi-soliton solution moves in space at the velocity ($v < c$) of the multi-soliton solution. Like the moving solution, $u$, $R[u]$ is also a Lorentz scalar. Moving solutions are obtained from static (1+2)-dimensional solutions by Lorentz transformations. Hence, it suffices to study the static solutions. Fig. 1 shows the vertex map of a static two-soliton solution.

When $V(q^{(1)}, q^{(2)}) < 0$, $R[u]$ is singular along a line in the $\{\xi_1$-$\xi_2\}$ plane. This corresponds to soliton solutions that are faster-than-light, which are not discussed in this paper. Moreover, the two subsets of solutions are not connected.

## 4. Particle interpretation of localized structures
### 4.1 Source-driven wave equation

The localized structure, $R[u]$ of Eq. (22), is *not* a solution of the Sine-Gordon equation but has its own equation of motion. It turns out to be easier to find this equation in the case of static multi-soliton solutions. Being independent of time (they are constructed through Eqs. (4)-(10), with the momentum vectors of Eq. (15)), static solutions obey the time-independent Sine-Gordon equation:

$$-\partial_x^2 u^{(S)}(x,y) - \partial_y^2 u^{(S)}(x,y) + \sin u^{(S)}(x,y) = 0 \quad . \tag{27}$$

In Eq. (27), $u^{(S)}(x,y)$ denotes the time-independent static soliton solution. Hence, $R[u^{(S)}]$ is the static version of the localized structure defined by Eq. (22).

Employing Eq. (27) repeatedly, one obtains the following equation for $R[u^{(S)}]$:

$$-\partial_x^2 R\left[u^{(S)}\right] - \partial_y^2 R\left[u^{(S)}\right] = 2\left(\left(u_{xy}^{(S)}\right)^2 - u_{xx}^{(S)} u_{yy}^{(S)}\right) \quad . \tag{28}$$

Eq. (28) for $R[u^{(S)}]$ is the Laplace equation in two space dimensions, driven by a source term. This source term is constructed from the static multi-soliton solution.

To obtain the equivalent of Eq. (28) in a moving frame, where the static solution, $u^{(S)}$, is replaced by a moving solution, $u$, one has to apply a Lorentz transformation to the coordinate vector, $x$ and to the momentum vectors, $q^{(i)}$. The transformation yields the relativistically invariant form of the wave equation for the l.h.s. of Eq. (28). However, when expressed in terms of $t$, $x$ and $y$, the source term on the r.h.s. assumes a complicated expression, which is not instructive. This can be overcome by exploiting the fact that all the momentum vectors, from which a multi-soliton solution is constructed, obey Eq. (13). Hence, a multi-soliton solution with $N \geq 2$ solitons depends only on two Lorentz scalars (together with the many $\alpha$- and $\beta$-coefficients):

$$\xi_i = q_\mu^{(i)} x^\mu \quad (i = 1,2) \quad . \tag{29}$$

In the static solution, one has (see Eq. (15))

$$\xi_i = \tilde{q}^{(i)}_\mu x^\mu = -\cos\psi^{(i)} x - \sin\psi^{(i)} y \qquad (i=1,2) \ . \tag{30}$$

In terms of these Lorentz scalars, the driving term in Eq. (28) can be re-expressed in a relativistically invariant form:

$$2\left(\left(u^{(S)}_{xy}\right)^2 - u^{(S)}_{xx} u^{(S)}_{yy}\right) = 2\left(1 - \left(\tilde{q}^{(1)} \cdot \tilde{q}^{(2)}\right)^2\right)\left(\left(u^{(S)}_{\xi_1 \xi_2}\right)^2 - u^{(S)}_{\xi_1 \xi_1} u^{(S)}_{\xi_2 \xi_2}\right)^2 \ . \tag{31}$$

The resulting equation for $R[u]$ in a moving frame becomes

$$\partial_\mu \partial^\mu R[u] = 2\left(1 - \left(q^{(1)} \cdot q^{(2)}\right)^2\right)\left(\left(u_{\xi_1 \xi_2}\right)^2 - u_{\xi_1 \xi_1} u_{\xi_2 \xi_2}\right) \ . \tag{32}$$

Eq. (32) can be verified directly by repeated application of Eqs. (3) and (9) to $R[u]$ of Eq. (22) in any Lorentz frame. Thus, the localized structure, obeys the wave equation, driven by a Lorentz invariant source term in any moving frame.

As expected, the source term vanishes identically on a single-soliton solution. The latter is obtained in the limit when all momenta become identical, so that both multiplicative factors on the r.h.s. of Eq. (32) vanish. Unlike $R[u]$, the source term need not be positive definite. Again, it suffices to show it in the rest frame, i.e., for Eq. (28). This is demonstrated in Fig. 3.

**4.2 Particle interpretation and mass generation**
$R[u]$ of Eq. (22) looks similar to the Lagrangian- or the Hamiltonian-densities associated with the Sine-Gordon equation, but coincides with neither. The fact that $R[u]$ is positive definite and spatially localized, suggests that one interpret it as a mass density of a spatially extended particle. In relativistic physics, the wave equation is associated with massless particles. Usually, massive particles are solutions of the wave equation, to which a mass term is added. The most common example is the Klein-Gordon equation,

$$\partial_\mu \partial^\mu w + m_0^2 w = 0 \ . \tag{33}$$

In the present case, of slower-than-light soliton solutions of the Sine-Gordon equation, the mass term is replaced by the soliton-driven source term on the r.h.s. of Eq. (32).

To complete the particle analogy, one has to assign the structure a mass. With the proposed interpretation, the mass is given as an integral of $R[u(x)]$ over space:

$$m = \int R[u(x)] d^2\vec{x} \ . \tag{34}$$

When $u(x)$ is a static soliton solution, it is at rest, hence and so is its vertex map, $R[u]$. Let us denote the rest mass of a vertex map by $m_0$. As pointed out previously, the Hirota solutions of Eq. (3) are Lorentz scalars. Hence, $R[u(x)]$ is also a Lorentz scalar. Consequently, under a Lorentz transformation to a moving frame, the mass of the "particle" generated from the moving solution through Eq. (32) will differ from the rest mass, only by the Jacobian of the space part of the Lorentz transformation, which is just the Lorentz factor, $\gamma$:

$$m = m_0 \gamma \qquad \left(\gamma = \left(1/\sqrt{1-v^2}\right)\right) \ . \tag{35}$$

The definition of the mass of localized structures as the space-integral of structure profile, as in Eq. (34), is commonly adopted in soliton dynamics. Usually there is no physical argument that leads to that definition (except in the case of solitons on the surface of an incompressible fluid, e.g., soliton solutions of the KdV equation). Here, this choice leads to Eq. (35).

### 4.3 "Bound states"?

An interesting observation emerges in solutions with $N \geq 3$ solitons. When all the free phases in Eq. (8) are small, there is only one soliton junction. In a static solution, it is localized around the origin in the $x$-$y$ plane. If, however, the phases are sufficiently sizable, then up to $N(N-1)/2$ distinct junctions (the maximal number of intersection points of $N$ lines in the plane) may exist. $R[u]$ of Eq. (22) then generates $N(N-1)/2$ distinct, spatially localized structures. Fig. 2 shows a vertex map of a static 3-soliton solution, with constant free phases so chosen that there are three distinct vertices. In a moving frame, all three move rigidly together at the same velocity, preserving their profiles. Moreover, the mass of the *whole* set of three vertices obeys Eq. (35). Thus, the three-some emulates a free, spatially extended particle. This is suggestive of a primitive emulation of a "bound state".

### 5. Lagrangian system

The coupling between the Sine-Gordon and wave equations in (1+2) dimensions, leading to Eq. (32), can be obtained from the-Euler-Lagrange equations of a system of two dynamical variables described by the following Lagrangian density:

$$L = \frac{1}{2}\partial_\mu w \partial^\mu w - (1-\cos w) + \frac{1}{2}\partial_\mu \rho \partial^\mu \rho + \varepsilon \rho J[w] \qquad (36)$$
$$(|\varepsilon| \ll 1, \ \mu = 0,1,2)$$

$L$ is the sum of the Lagrangian densities of the Sine-Gordon and the linear wave equations, and a small coupling term. The Euler-Lagrange equations are:

$$\partial_\mu \partial^\mu w + \sin w = \varepsilon \rho \frac{\delta J[w]}{\delta w} , \qquad (37)$$

$$\partial_\mu \partial^\mu \rho = \varepsilon J[w] . \qquad (38)$$

Here $\delta$ denotes a variational derivative. Let us expand $w$ and $\rho$ in powers of $\varepsilon$:

$$w = u + \varepsilon u^{(1)} + \varepsilon^2 u^{(2)} + ... , \qquad (39)$$

$$\rho = \rho^{(0)} + \varepsilon \rho^{(1)} + \varepsilon^2 \rho^{(2)} ... . \qquad (40)$$

Inserting Eqs. (39) and (40) in Eqs. (37) and (38), the order-by-order equations through $O(\varepsilon)$ are:

$$\partial_\mu \partial^\mu u + \sin u = 0 , \qquad (41)$$
$$\partial_\mu \partial^\mu \rho^{(0)} = 0 , \qquad (42)$$
$$\partial_\mu \partial^\mu u^{(1)} + u^{(1)} \cos u = \rho^{(0)} \frac{\delta J[u]}{\delta u} , \qquad (43)$$
$$\partial_\mu \partial^\mu \rho^{(1)} = J[u] . \qquad (44)$$

Eq. (41) is the (1+2)-dimensional Sine-Gordon equation, and Eq. (42) is the linear wave equation, which has the capacity to generate massless particles (photons). To focus on slower-than-light solutions, one must preclude the possibility of massless solutions, for which there is no rest frame. To this end, let us omit the $O(\varepsilon^0)$-term in Eq. (40) and focus on solutions, in which $\rho$ is of $O(\varepsilon)$:

$$\rho^{(0)} = 0 \ . \tag{45}$$

Eq. (43) is then homogeneous (its r.h.s. vanishes), allowing for

$$u^{(1)} = 0 \ . \tag{46}$$

For solutions obeying Eqs. (45) and (46), the $O(\varepsilon^2)$ equations are found to be:

$$\partial_\mu \partial^\mu u^{(2)} + u^{(2)} \cos u = \rho^{(1)} \frac{\delta J[u]}{\delta u} \tag{47}$$

and

$$\partial_\mu \partial^\mu \rho^{(2)} = 0 \ . \tag{48}$$

Thus, again, to preclude the possibility of massless solutions, for which there is no rest frame, let us focus on the solution, for which

$$\rho^{(2)} = 0 \ . \tag{49}$$

In summary, the order-by-order expansion of the Euler-Lagrange equations of the system described by the Lagrangian of Eq. (36) admits a solution, which, in $O(\varepsilon^0)$, is a soliton solution of the (1+2)-dimensional Sine-Gordon equation, and, the $O(\varepsilon)$ part is a solution of Eq. (44), the linear wave equation driven by a source-term that is constructed out of soliton solutions of Eq. (41). The effect of the dynamical variable, $\rho$, on Sine-Gordon solitons occurs only in $O(\varepsilon^2)$.

All that remains is to choose the source term, so that Eq. (44) becomes Eq. (32), with $\rho^{(1)}$ playing the role of $R[u]$ of Eq. (22). Obtaining the desired result is easiest in the rest frame, where Eq. (28) holds. The driving term in Eq. (28) is generated by the following Lagrangian density:

$$L = -\frac{1}{2}\left((\partial_x w)^2 + (\partial_y w)^2\right) - (1 - \cos w) - \frac{1}{2}\left((\partial_x \rho)^2 + (\partial_y \rho)^2\right) + 2\varepsilon \rho\left((w_{xy})^2 - w_{xx} w_{yy}\right) \ . \tag{50}$$

Eq. (44) for $\rho^{(1)}$ becomes Eq. (28) for $R[u]$ in the rest frame, and Eq. (32) – in a moving frame.

The $O(\varepsilon^2)$-equations obtained from Eqs. (47) and (48) are:

$$-u^{(2)}_{xx} - u^{(2)}_{yy} + \cos u \, u^{(2)} = 2\left(R_{xy} u_{xy} - R_{xx} u_{yy} - R_{yy} u_{xx}\right) . \tag{51}$$

$$-\rho^{(2)}_{xx} - \rho^{(2)}_{yy} = 0 , \tag{52}$$

Thus, the massive, spatially extended particle-like structure affects Sine-Gordon solitons only in $O(\varepsilon^2)$. Finally, in the absence of any interaction term in Eq. (52), the simplest choice is $\rho^{(2)} = 0$. (this precludes massless particles in the moving frame.)

## 6. Concluding comments
Several comments are due here.

1) Being generated out of multi-soliton solutions, the structures that emulate free, spatially extended, massive relativistic particles may be viewed as composite "particles".

2) The end result may be viewed as a system of two coupled equations, for two dynamical variables, one yielding the Sine-Gordon solitons, the other - a localized entity, which may be interpreted as the mass density of a free, spatially extended relativistic particle.

3) The resulting system can be described in terms of a (1+2)-dimensional dynamical system that involves two degrees of freedom, with a Lagrangian density, in which the Sine-Gordon equation and the linear wave equation are coupled by a weak coupling term.

4) This idea is applicable to other relativistically invariant evolution equations, which have multi-soliton solutions. A single-soliton identity, the analog of the vanishing of $R$ of Eq. (22) on the single-Sine-Gordon soliton, needs to be derived. The resulting functional automatically generates structures that are localized around soliton junctions when computed for multi-soliton solutions. These structures emulate spatially extended particles. If the equation of motion of their mass density is found, it leads to the desired Lagrangian density.

**Appendix I: Lorentz transformation**

A single-soliton solution is constructed in terms of one momentum vector. In solutions with $N \geq 3$, only two of the momentum vectors are independent. Hence, it suffices to discuss the properties of one and two tachyonic momentum vectors under Lorentz transformations.

A Lorentz transformation in (1+2) dimensions is given by the following matrix:

$$L = \begin{pmatrix} \gamma & -\gamma \beta_x & -\gamma \beta_y \\ -\gamma \beta_x & 1+(\gamma-1)\frac{\beta_x^2}{\beta^2} & (\gamma-1)\frac{\beta_x \beta_y}{\beta^2} \\ -\gamma \beta_y & (\gamma-1)\frac{\beta_x \beta_y}{\beta^2} & 1+(\gamma-1)\frac{\beta_y^2}{\beta^2} \end{pmatrix}. \quad (I.1)$$

In Eq. (I.1)

$$\vec{\beta} = \{\beta_x, \beta_y\} = \{v_x, v_y\}/c \quad , \quad \gamma = 1/\sqrt{1-\vec{\beta}^2} \quad , \quad (I.2)$$

where $v_x$ and $v_y$ are the components of the velocity of the Lorentz transformation. A single (1+2)-dimensional vector, which obeys Eq. (9), can be transformed into a vector of the form given in Eq. (15) by a one-parameter family of transformations, as there are two free parameters, $\beta_x$ and $\beta_y$.

The situation is different when two vectors $q^{(1)}$ and $q^{(2)}$, which obey Eq. (9), are considered. Applying the transformation of Eq. (I.1) to the two vectors

$$q^{(i)} = \{q_0^{(i)}, q_x^{(i)}, q_y^{(i)}\} \quad (i=1,2) \quad , \quad (I.3)$$

one obtains the expressions for $\beta_x$ and $\beta_y$, which are required for the transformed vectors to have vanishing time components, as in Eq. (15):

$$\beta_x = \frac{q_0^{(1)} q_y^{(2)} - q_0^{(2)} q_y^{(1)}}{q_x^{(1)} q_y^{(2)} - q_x^{(2)} q_y^{(1)}} \quad , \quad \beta_y = \frac{q_0^{(2)} q_x^{(1)} - q_0^{(1)} q_x^{(2)}}{q_x^{(1)} q_y^{(2)} - q_x^{(2)} q_y^{(1)}} \quad . \quad (I.4)$$

For the transformation to be feasible, its velocity must be lower than $c$. Hence, the magnitude of the vector $\vec{\beta}$ must be smaller than 1. Using Eq. (I.4), one obtains the constraint:

$$1-\beta_x^2-\beta_y^2 = \frac{1-\left(q^{(1)}\cdot q^{(2)}\right)^2}{\left(q_x^{(1)}q_y^{(2)}-q_x^{(2)}q_y^{(1)}\right)^2} > 0 \quad . \tag{I.5}$$

Thus, for a pair of vectors that obey Eq. (9) to be transformable to the form given in Eq. (15), its scalar product in Minkowski space must obey

$$\left|q^{(1)}\cdot q^{(2)}\right| < 1 \quad . \tag{I.6}$$

The pair of solitons that are constructed from such two vectors moves rigidly at a velocity, given in Eq. (I.4), that is lower than $c$. This pair of solitons is transformed by the Lorentz transformation to a rest frame, in which both solitons are static. If a solution contains $N \geq 3$ solitons, Eq. (13) guarantees that the solution propagates rigidly as a whole at a velocity that is lower than $c$.

If the inequality is inverted,

$$\left|q^{(1)}\cdot q^{(2)}\right| \geq 1 \quad , \tag{I.7}$$

then there is no velocity lower than $c$ that can yield the desired transformation. The pair of solitons generated from such two vectors propagates at a velocity, $v \geq c$.

The limit of equality,

$$q^{(1)}\cdot q^{(2)} = \pm 1 \quad , \tag{I.8}$$

cannot be reached from within the family of slower-than-light solutions. An $N$-soliton solution, which propagates at a velocity $v < c$, degenerates into a solution with a smaller number of solitons: ($N$–1) solitons in the case of the (+) sign, and ($N$–2) solitons in the case of the (–) sign. These solutions are also slower-than-light. Hence, one cannot reach the speed of light from below. However, the limit of Eq. (I.8) can be achieved within the family of solutions that obey Eq. (I.7). The two solution subsets are not connected.

Acknowledgments: The author wishes to thank G. Bel and G. Burde for constructive discussions.

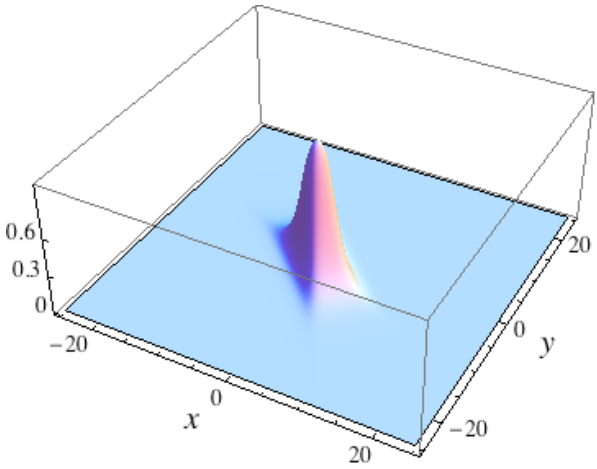

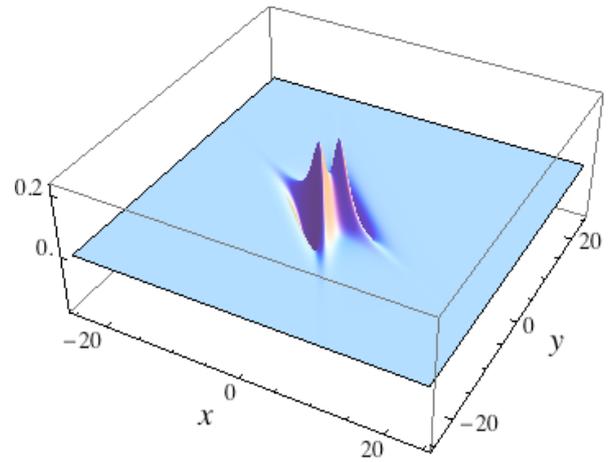

Fig. 1 Vertex map of static two-soliton solution in (1+2) dimensions.

Momenta (Eq. (13)): $\psi^{(1)} = \pi/6$; $\psi^{(2)} = \pi/3$; Phase shifts (Eq. (8)): $\delta_1 = \delta_2 = 0$.

Fig. 3 Source term in Eq. (29) constructed from static two-soliton solution. Parameters as in Fig. 1

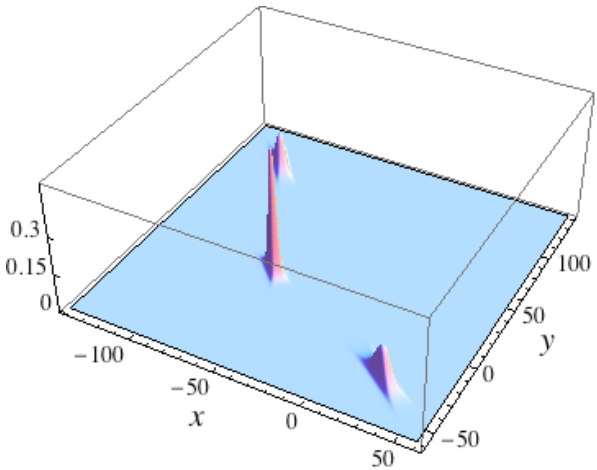

Fig. 2 Vertex map of static three-soliton solution in (1+2) dimensions with phase-

shifted solitons. Momenta (Eq. (13)): $\psi^{(1)} = \pi/6$; $\psi^{(2)} = \pi/3$; $\psi^{(3)} = \pi/4$;

Phase shifts (Eq. (8)): $\delta_1 = -40$; $\delta_2 = -10$; $\delta(3) = 0$.